\documentclass{ws-ijmpe}
\usepackage[super,compress]{cite}
\usepackage{color}
\usepackage[utf8]{inputenc}
\usepackage[english]{babel}
\usepackage{hyperref}
\begin{document}
\markboth{M. Kumawat \textit{et al.}}{Novel feature of doubly bubble nuclei in 50$\leq$Z(N)$\leq$82 region}
\catchline{}{}{}{}{}
\title{\textsc{Novel feature of doubly bubble nuclei in 50$\leq$Z(N)$\leq$82 region along with magicity and weakly bound structure}}

\author{M. Kumawat}
\address{Department of Physics, School of Basic Sciences, Manipal University Jaipur, Jaipur-303007, India}
\author{G. Saxena}
\address{Department of Physics (H $\&$ S), Govt. Women Engineering College, Ajmer-305002, India\\ gauravphy@gmail.com}
\author{M. Kaushik}
\address{S. S. Jain Subodh P. G. College, M. C. A. Institute, Rambagh Circle, Jaipur-302004, India}
\author{S. K. Jain}
\address{Department of Physics, School of Basic Sciences, Manipal University Jaipur, Jaipur-303007, India}
\author{J. K. Deegwal}
\address{Govt. Women Engineering College, Ajmer-305002, India}
\author{Mamta Aggarwal}
\address{Department of Physics, University of Mumbai, Kalina Campus, Mumbai-400098, India\\
mamta.a4@gmail.com}

\maketitle

\begin{history}
\received{Day Month Year}
\revised{Day Month Year}
\end{history}

\begin{abstract}
In this work, we identify a unique and novel feature of central density depletion in both proton and neutron named as doubly bubble nuclei in 50$\leq$Z(N)$\leq$82 region. The major role of 2d-3s single-particle (s.p.) states in the existence of halo and bubble nuclei is probed. The occupancy in s.p. state 3s$_{1/2}$ leads to the extended neutron density distribution or halo while the unoccupancy results in the central density depletion. By employing the Relativistic Mean-Field (RMF) approach along with NL3* parameter, the separation energies, single-particle energies, pairing energies, proton, and neutron density profiles along with deformations of even-even nuclei are investigated. Our results are in concise with few other theories and available experimental data. Emergence on new shell closure and the magicity of conventional shell closures are explored systematically in this yet unknown region.
\end{abstract}

\keywords{ Relativistic mean-field plus BCS approach; 50$\leq$Z(N)$\leq$82; Shell Closure; Bubble Structure; Weakly bound nuclei.}



\section{Introduction}
Remarkable advancements of various experimental facilities and salient development in accelerator
and detection technology during the last three decades have prompted the interest to examine various exotic properties viz. new shell closure, weakly bound structure or halo, bubble structure, etc. in the obscure domain of periodic chart~\cite{tanihata,kanungo1,ozawa,ozawa1}. As a result, confirmation of new magicity in sd-shell at neutron number N$=$14~\cite{stanoiu,brown,becheva}, N$=$16~\cite{hoffman,tshoo,kanungo} and proton number Z$=$16~\cite{togano1} as well as in pf-shell at N$=$32~\cite{gade,wienholtz,rosenbusch} and N$=$34~\cite{stepp} is observed. Interestingly, the break down of conventional magicity N$=$8, 20, 28, etc. is also reported~\cite{iwasaki,door,bastin} by various experimental investigations. Besides these exotic features, the existence of extended density distribution in $^{22}$C is also observed by recent reaction cross-section measurements~\cite{tanaka}. Indeed, a halo structure is suggested due to the last two neutrons predominantly in the 2s$_{1/2}$ orbit for $^{22}$C with very weak binding~\cite{gaudefroy,kobayashi}. After recent interaction cross-section study by Togano \textit{et al.} and using low-energy limit of neutron-neutron interaction by Suzuki \textit{et al.}~\cite{suzuki}, halo structure in $^{22}$C~\cite{togano} and $^{24}$O have been confirmed. The role of 2s$_{1/2}$ orbit is found crucial as well indicating proton bubble structure in $^{34}$Si~\cite{nature} in the sd shell region. Recently, these important features in sd-shell nuclei and pf-shell nuclei are demonstrated using the relativistic mean-field approach~\cite{saxena,saxena1}.\par

In anticipation with the similar characteristics of 3s$_{1/2}$ orbit, as mentioned above for 2s$_{1/2}$ orbit, the present theoretical study for nuclei between 50$\leq$Z(N)$\leq$82 which are mainly governed by 2d-3s single-particle states is proposed and aimed. To the best of our knowledge, this kind of systematic study of these nuclei covering nuclei between 50$\leq$Z(N)$\leq$82 considering exotic phenomenon like new magicity, the disappearance of conventional magicity, weakly bound structure or halo and with doubly central depletion as well. However, it is worthy to mention here that proton bubble is indeed visible in $^{206}$Hg due to substantial role of 3s$_{1/2}$ orbit as per relativistic mean-field (RMF) approach \cite{todd}. It has also been shown that the semibubble structure in $^{204,206}$Hg persists not only in the ground state but also in their excited states \cite{wu2018}. On the other side, the phenomena of giant halo due to filling in of 3s$_{1/2}$ orbit in neutron-rich even-Ca isotopes is already investigated by relativistic continuum Hartree-Bogoliubov (RCHB), non-relativistic Skyrme Hartree-Fock-Bogoliubov (HFB) calculations~\cite{terasaki} and also by RMF+BCS approach using TMA and NL-SH parameter~\cite{kaushik}.\par

This particular region is also very important for the astrophysical purpose and there have been few studies which target mainly the astrophysical implication of these nuclei. For example, the optical potential was generated folding the nuclear density with the microscopic nuclear interaction DDM3Y to study low-energy proton reactions for different nuclei in the A $\approx$ 40-120 region~\cite{lahiri2011,lahiri2012,dutta2015,lahiri2016}. In addition, the mass region ranging from A $\approx$ 74 to 196 is explored to calculate astrophysical S-factor for 36 known p-nuclei with (p, $\gamma$) reactions at low energy taking spherical densities from RMF calculations~\cite{naik2018}. On the other side, a number of studies have been completed~\cite{watanabe2013,simpsond2014,taprogge2014} in the neighborhood of the N$=$82, A$=$130 r-process peak. In particular, the first-ever study ~\cite{watanabe2013} of the level structure of the waiting-point nucleus $^{128}$Pd and $^{126}$Pd has been indicated that the shell closure at the neutron number N$=$82 is fairly robust.\par

Looking towards the importance of this region between 50$\leq$Z(N)$\leq$82 which exhibits varieties of exotic features and still scarce with theoretical and systematic treatment, has invoked this present theoretical study. Therefore, in this communication, we describe ground-state properties of even-even nuclei in which neutron or proton lies between Z(N)$=$50 and 82. To probe magicity, weakly bound structure, and central depletion (bubble structure) we examine various properties viz. binding energies, single-particle energies, deformations, separation energies, pairing energies as well as density distributions, etc. For our calculations, we use Relativistic Mean-Field (RMF) approach~\cite{walecka,boguta,bouy,pgr2,suga,ring,sharma,lala,mizu,estal,yadav,yadav1,vretenar2005,meng2006,niksic2011,meng2016} with NL3* parameter~\cite{nl3star}. We also compare our results along with available experimental data~\cite{nndc} and other theories to testify our outcomes.

\section{Theoretical Frameworks}
We employ relativistic mean-field approach~\cite{saxena,saxena1,walecka,boguta,bouy,pgr2,suga,ring,sharma,lala,mizu,estal,yadav,yadav1,saxena2} together with a realistic mean-field, which has proved to be very useful and a successful tool as shown in our earlier work ~\cite{saxena,saxena1,saxena2}. We use the model Lagrangian density with nonlinear terms both for the ${\sigma}$ and ${\omega}$ mesons as described in detail in Refs.~\cite{suga,yadav,singh}. For the pairing interaction, we use a delta force, i.e. V$=$-V$_0 \delta(r)$ with the strength V$_0$$=$350 MeV fm$^3$ same as has been used in Refs.~\cite{saxena,saxena1,yadav,saxena2} for the description of drip-line nuclei. Based on the single-particle spectrum calculated by the RMF, we perform a state dependent BCS calculations~\cite{lane,ring2}. This approach has proven to be very successful for recent extensive study of (i) conventional and new magic nuclei ~\cite{saxena,saxena1,yadav1,saxena2,saxena3,saxena4,saxena5,CJP}, (ii) describing interdependence of 2p-halo with 2p-radioactivity \cite{saxena6} and (iii) Bubble structure ~\cite{saxena,saxena7}.\par

Another formalism used for the present study is the triaxially deformed Nilson Strutinsky (NS) model which treats the delicate interplay of macroscopic bulk properties of nuclear matter and the microscopic shell effects and has been used extensively in our earlier works~\cite{MAMPRC,aggarwal}. We evaluate binding energy, separation energy, deformation, and shape by incorporating macroscopic binding energy BE$_{LDM}$ obtained from the LDM mass formula~\cite{PM} to the microscopic effects arising due to nonuniform distribution of nucleons through the Strutinsky's shell correction $\delta$E$_{shell}$~\cite{VM} along with the deformation energy E$_{def}$ obtained from the surface and Columb effects~\cite{MAMPRC}.\par

The detailed description of these theoretical formalisms that have been adequately described in our earlier works has not been given in this paper. (Readers may refer~\cite{suga,yadav,singh,meng5} for detailed description of RMF approach and~\cite{MAMPRC,aggarwal} for NSM formalism).\par

\section{Results and discussions}
\begin{figure}[!htbp]
\centering
\includegraphics[width=0.8\textwidth]{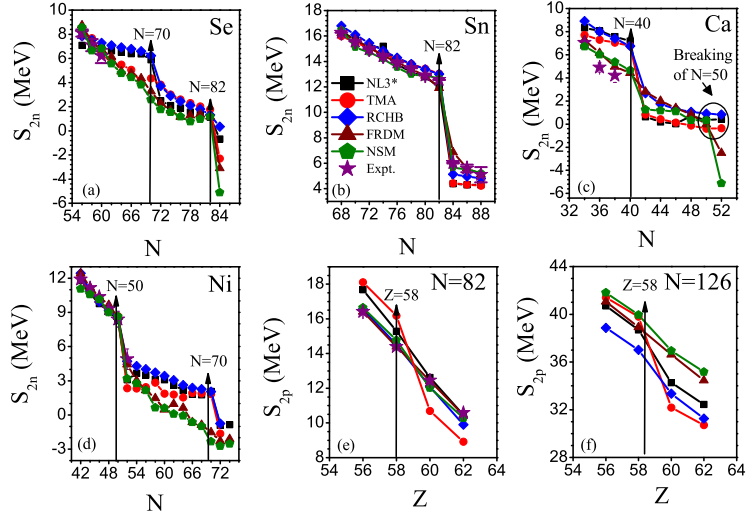}
\caption{(Colour online) Two neutron separation energy (S$_{2n}$) and two proton separation energy (S$_{2p}$) of various isotopes/isotones.} \label{fig1}
\end{figure}

The 2d-3s shell in the 50$\leq$Z(N)$\leq$82 region plays a major role in the formation of halo and bubble like structures and even influences the magicity. Since these exotic phenomena of halo, bubble and magicity at drip lines have not been much explored in this region, we make a systematic study in this work. As per shell model convention, 2d-3s shell consists of single-particle states 2d$_{5/2}$, 3s$_{1/2}$ and 2d$_{3/2}$ sandwiched between 1g$_{7/2}$ and 1h$_{11/2}$ states. This region consisted of two conventional magic numbers 50 and 82 whereas two new magic numbers 58 and 70 have been speculated to exist \cite{nica2007,nakada2014,madhu2005}. For instance, the measured first excitation energy is found slightly higher in $^{140}$Ce than in the surrounding N$=$82 isotones $^{138}$Ba and $^{142}$Nd \cite{nica2007} which is one of the indicator of magic character of Z$=$58, which has been shown in another work to have a sub-magic character at $^{140}$Ce by self-consistent mean-field calculations with the M3Y-P6 interaction \cite{nakada2014}. A new magic number has been reported at N$=$70 using RMF formalism in the coordinate space \cite{madhu2005} in Ni nuclei.\par

To explore the possibility of these new magic numbers, we have plotted two neutron separation energy (S$_{2n}$) and two proton separation energy (S$_{2p}$) of a few selected isotopes/isotones in Fig. \ref{fig1}. These energies are calculated using the NL3* parameter~\cite{nl3star} of the RMF approach and compared with the similar RMF calculations done with TMA parameter~\cite{suga-tma}. Moreover, we have also calculated S$_{2n}$ and S$_{2p}$ using the NSM approach~\cite{MAMPRC,aggarwal} which is found in a very good match with our RMF calculations. We also compare data with that of FRDM~\cite{frdm2012}, RCHB~\cite{rchb} along with experimental data~\cite{nndc} for the comparison. From Fig. \ref{fig1}(a) and (b) a sharp drop in S$_{2n}$ after N$=82$ is observed in Se and Sn isotopes which indicates strong magicity of N$=82$ in accord with the study by Watanabe \textit{et al.}~\cite{watanabe2013} which stated that the shell closure at the neutron number N$=$82 is fairly robust. Fig. \ref{fig1}(c) shows a sharp drop in S$_{2n}$ after N$=$40 in Ca isotopes which indicates a new magic number N$=$40 ($^{60}$Ca) at drip line whereas the similar steep decline in S$_{2n}$ value is missing in the conventional magic number N$=$50 which is a sign of breaking the conventional magicity of N$=$50 at the neutron drip-line. Such kind of appearance of new shell closure at N$=$40 and disappearance of conventional shell closure at N$=$50 is an important outcome for experimental studies such as the recent study done by Tarasov \textit{et al.} in which discovery of $^{60}$Ca is reported and $^{70}$Ca is anticipated to be a drip-line of Ca isotopes. However, magicity of N$=$50 remains strong in $^{78}$Ni (Fig. \ref{fig1}(d)) in accord with our earlier result \cite{saxena2}. Ni isotopes also exhibit magicity in N$=$70 with a sharp drop in S$_{2n}$ after N$=$70 in agreement with the results of the magicity of N$=$70 using RMF formalism in the coordinate space \cite{madhu2005}.  Also Se isotopes show significant magicity at N$=$70 (Fig. \ref{fig1}(a)). Another new magic number is predicted at Z$=$58 as seen in the plots of S$_{2p}$ vs. Z for N$=$82, 126 isotones in Fig. \ref{fig1}(e) and (f) inline with Refs. \cite{nica2007,nakada2014,madhu2005}. Therefore, Fig. \ref{fig1} shows the emergence of two new magic numbers Z$=$58, a new proton magic number and N$=$70, a new neutron magic number.\par

\begin{figure}[!htbp]
\centering
\includegraphics[width=0.6\textwidth]{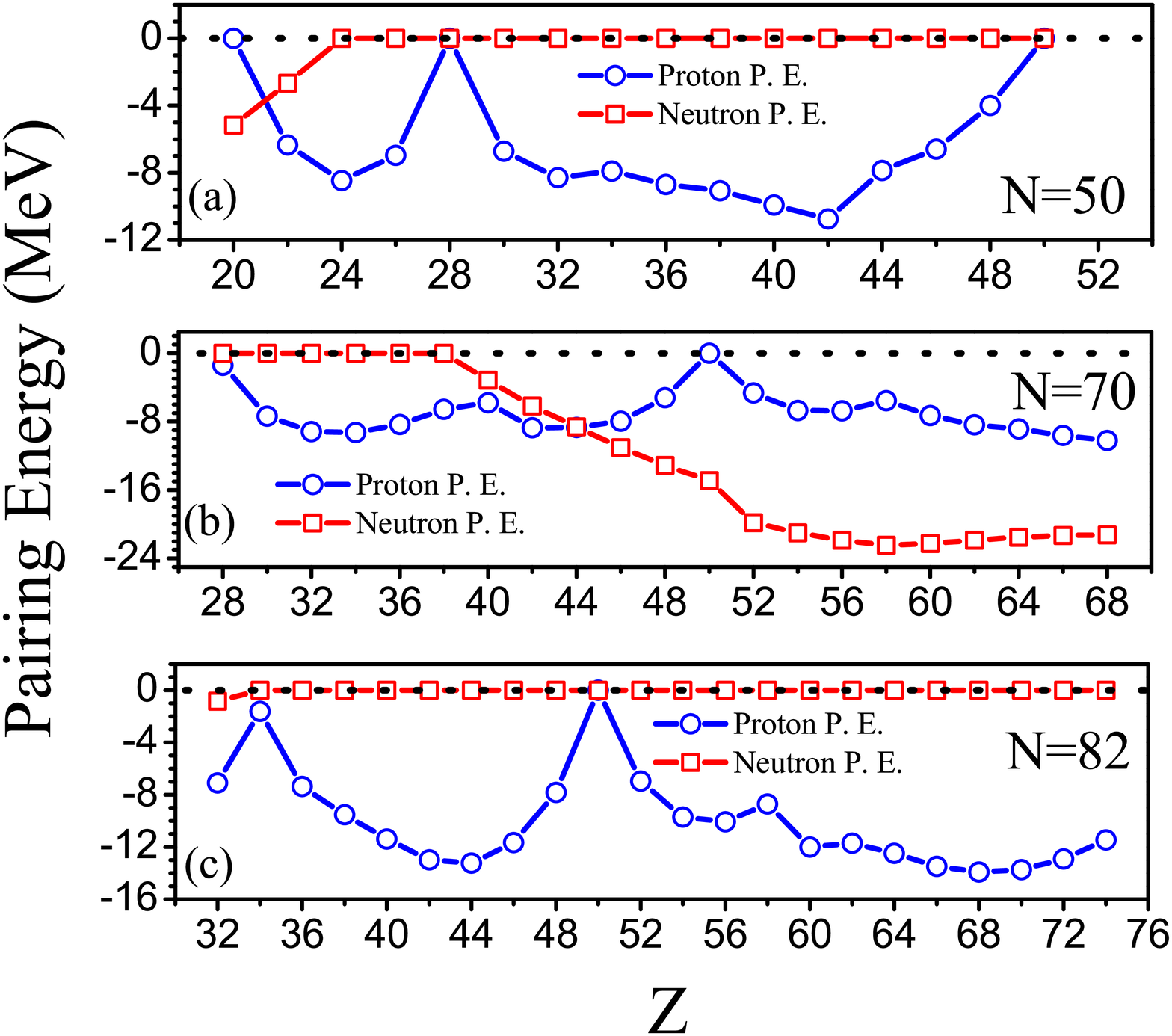}
\caption{(Colour online) Neutron and proton pairing energy of isotones with neutron number N$=$50, 70 and 82.} \label{fig2}
\end{figure}

It is well-known ~\cite{yadav,saxena4} that the pairing energy vanishes for the magic nuclei and therefore the analysis of pairing energy offers evidence for the existence of magicity or the shell closures. For this analysis, we show proton and neutron pairing energy in Fig. \ref{fig2} for the isotonic chain with N$=$50, 70, and 82. In the case of N$=$50, the magicity with zero neutron pairing is seen in Fig. \ref{fig2}(a) for mostly isotones except $^{70}$Ca and $^{72}$Ti, in which N$=$50 does not remain magic and shows the breakdown or weakening towards neutron drip-line in accordance with the study done by Yadav \textit{et al.} ~\cite{yadav}. Fig. \ref{fig2}(b) shows magic behaviour for N$=$70 for neutron-rich isotones with Z$=$28-38 where neutron pairing energy is zero. The neutron pairing energy is found zero for all the isotones for the case of N$=$82 as seen in Fig. \ref{fig2}(c) which points towards a very strong magicity of N$=$82 for the full isotonic chain. From proton pairing energy curves in all the panels, one can see that the proton pairing energy falls in between the two closed shells which indicate non-magic nuclei. Whenever proton pairing energy is zero along with zero neutron pairing energy, one gets a doubly magic nucleus. The pairing energy increases from zero at shell closure to a maximum value in the mid shell nuclei showing pairing interaction. The proton pairing energy vanishes for Z$=$20, 28 and 50 in all the plots and we find doubly magic nuclei $^{78}$Ni and $^{100}$Sn in N$=$50 isotones, $^{98}$Ni in N$=$70 isotones, and $^{132}$Sn in N$=$82 isotones. In addition, the kink due to lower pairing energy comparative to their neighbourhood isotopes/isotones e.g. at Z$=$34, Z$=$40, and Z$=$58 in Fig. \ref{fig2}(a), (b) and (c), respectively, hint towards the new sub-shell closure in this region.\par

\begin{figure}[!htbp]
\centering
\includegraphics[width=0.6\textwidth]{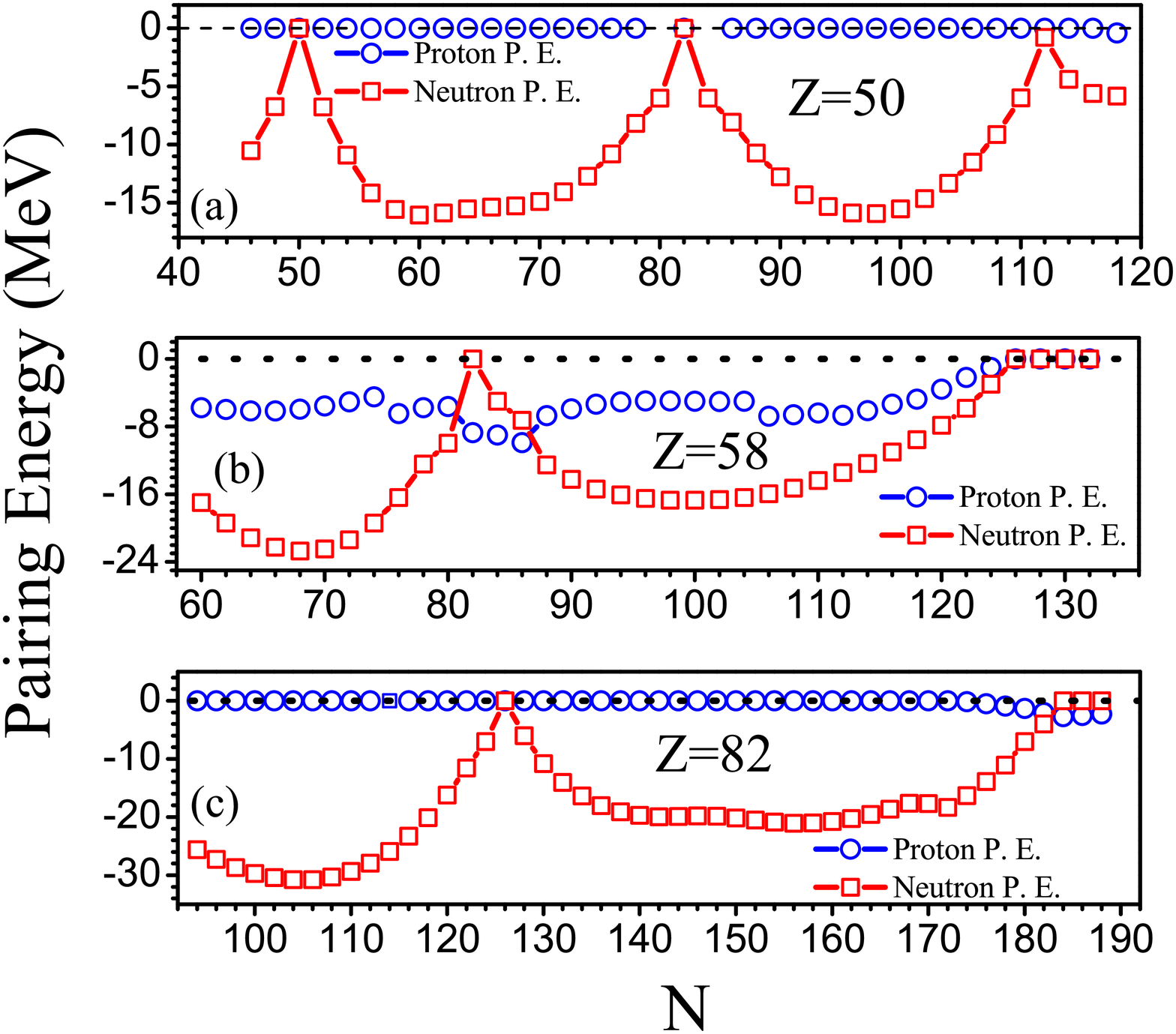}
\caption{(Colour online) Neutron and proton pairing energy of isotopes with proton number Z$=$50, 58 and 82.} \label{fig3}
\end{figure}

In Fig. \ref{fig3}, we have shown the pairing energy of the isotopic chains of Z$=$50, 58, and 82. The Z$=$50 and 82 (Sn and Pb) isotopes are found very strongly magic throughout the chain as can be seen from Fig. \ref{fig3}(a) and (c). The neutron pairing contribution of these isotopes shows $^{100, 132}$Sn and $^{208}$Pb as the strong doubly magic candidates with zero neutron and proton pairing energies. We also find new doubly magic nuclei $^{172}$Sn and $^{286}$Pb with the emerging new neutron magic number N$=$112 and 184, respectively. As far as the isotopic chain of Z$=$58 is concerned (shown in Fig. \ref{fig3}(b)), it does not show zero pairing energy for most of the isotopes unlike Sn and Pb isotopes. However, the proton pairing energy is either close to zero or towards the lower side for almost throughout the chain than the neutron pairing energy trend. This behaviour of Z$=$58 indicates this proton number as a sub-shell closure and shows a doubly magic character at N$=$126 in $^{184}$Ce as is seen in Fig. \ref{fig3}(b) where both pairing energy contributions vanish.\par

\begin{figure}[!htbp]
\centering
\includegraphics[width=0.6\textwidth]{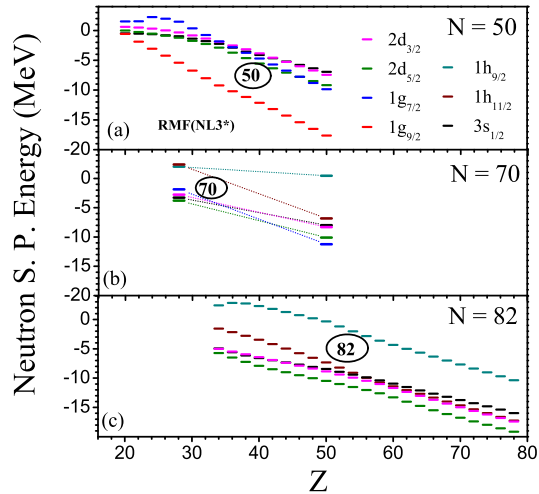}
\caption{(Colour online) Neutron single-particle energies of neutron number N$=$50, 70 and 82 isotones.} \label{fig4}
\end{figure}

In order to probe more into magicity, we display neutron single-particle states of isotones with neutron number N$=$50, 70, and 82 in Fig. \ref{fig4}(a), (b) and (c), respectively. For N$=$70, we have shown s.p. states of two isotones $^{98}$Ni and $^{120}$Sn which have been found to be spherical in our calculations. In Fig. \ref{fig4}(a), we observe a significant energy gap between 1g$_{9/2}$ and 1g$_{7/2}$ states leading to N$=$50 shell closure for proton-rich doubly magic $^{100}$Sn. But this gap decreases from 7 MeV (at $^{100}$Sn) to 2 MeV (at $^{70}$Ca) as one moves towards proton deficient side. Therefore, the N$=$50 shell closure weakens for the nuclei with Z $\approx$ 20 ($^{70}$Ca, $^{72}$Ti, etc.) as is also indicated by the pairing energy systematics (Fig. \ref{fig2}(a)). On the other side, in Fig. \ref{fig4}(b), the development of a new single-particle gap and hence a new shell closure N$=$70 is observed while moving towards proton-rich N$=$70 isotones. This gap arises between the 2d-3s shell and 1h$_{11/2}$ state which goes to a maximum value of 5.6 MeV for Z$=$28 leading to doubly magic nucleus $^{98}$Ni. For N$=$82 isotones, from Fig. \ref{fig4}(c), it is clear that the gap between 1h$_{11/2}$ and 1h$_{9/2}$ remains significantly large around 4-7 MeV ensuing N$=$82 with strong magic character.\par

\begin{figure}[!htbp]
\centering
\includegraphics[width=0.6\textwidth]{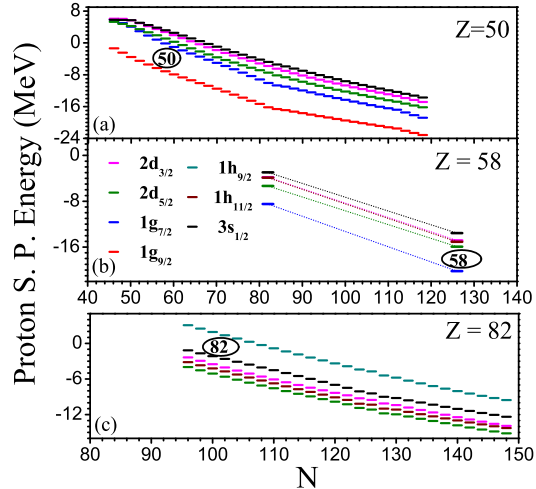}
\caption{(Colour online) Proton single-particle energies of proton number Z$=$50, 58 and 82 isotopes.} \label{fig5}
\end{figure}

To look for the proton magicity in proton 2d-3s shell nuclei, we show proton single-particle states of Z$=$50, 58, and 82 (Sn, Ce, and Pb) isotopes in Fig. \ref{fig5}(a), (b) and (c), respectively. Fig. \ref{fig5}(a) shows the complete chain of Sn isotopes where the gap between proton 1g$_{9/2}$ and 1g$_{7/2}$ states persists upto a value of $\sim$ 5 MeV which reassures the strong magicity of Z$=$50. A small gap of around ($\geq$ 3 MeV) is observed between 1g$_{7/2}$ and 2d-3s shell which may give rise to (sub)shell closure at Z$=$58. Here we have shown two isotopes $^{140}$Ce and $^{184}$Ce from the Z$=$58 isotopic chains which are found spherical in our calculations in accord with almost zero deformation in them as predicted by our calculations using NSM \cite{MAMPRC,aggarwal} and data from FRDM \cite{frdm2012}, HFB \cite{goriely} and WS4 \cite{ws42014} available experimental data \cite{nndc}. These two isotopes show the magic character among Z$=$58 isotopes in concise with \cite{nica2007,nakada2014}. From Fig. \ref{fig5}(c), the strong magicity of Pb isotopes (Z$=$82) is clearly visible and hence we ascertain the strong magic character of N and Z$=$82.\par

\begin{figure}[!htbp]
\centering
\includegraphics[width=0.6\textwidth]{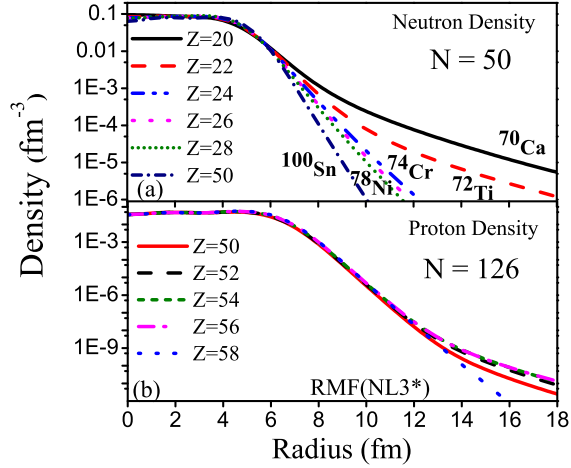}
\caption{(Colour online) Variation of (a) neutron density for N$=$50 isotones and (b) proton density for N$=$126 isotones with respect to radius.} \label{fig6}
\end{figure}

To get more insight into the weakening of magicity at N$=$50 and the evolution of the new magic number at Z$=$58, we display a density distribution of N$=$50 and N$=$126 isotones. One can see from Fig. \ref{fig6}(a) that while moving from the neutron deficient nuclei $^{100}$Sn (Z$=$50) towards neutron richer side $^{74}$Cr (Z$=$24) and then to very neutron-rich $^{70}$Ca, neutron density distributions show sharp fall for neutron deficient $^{100}$Sn, $^{78}$Ni, up to $^{74}$Cr showing the confinement of the distribution up to smaller distances which indicate the neutron magic character of these isotones which gradually starts extending to longer tails for $^{70}$Ca and $^{72}$Ti. This extended distribution at $^{70}$Ca and $^{72}$Ti nuclei prove the disappearance of N$=$50 shell closure at the neutron drip line side. As mentioned above, this observation of the disappearance of shell closure at $^{70}$Ca be very useful and important that can also provide an additional impulse for further experimental study after the recent discovery of $^{60}$Ca and interpretation on $^{70}$Ca \cite{tarasov}. From Fig. \ref{fig6}(b), interestingly, the proton density distribution of Z$=$58 is found to falls off at a smaller distance as compared to the other nuclei even with smaller Z. This characteristic brings in the sub-shell character of Z$=$58 in the nuclei of 2d-3s shell.\par

\begin{figure}[!htbp]
\centering
\includegraphics[width=0.6\textwidth]{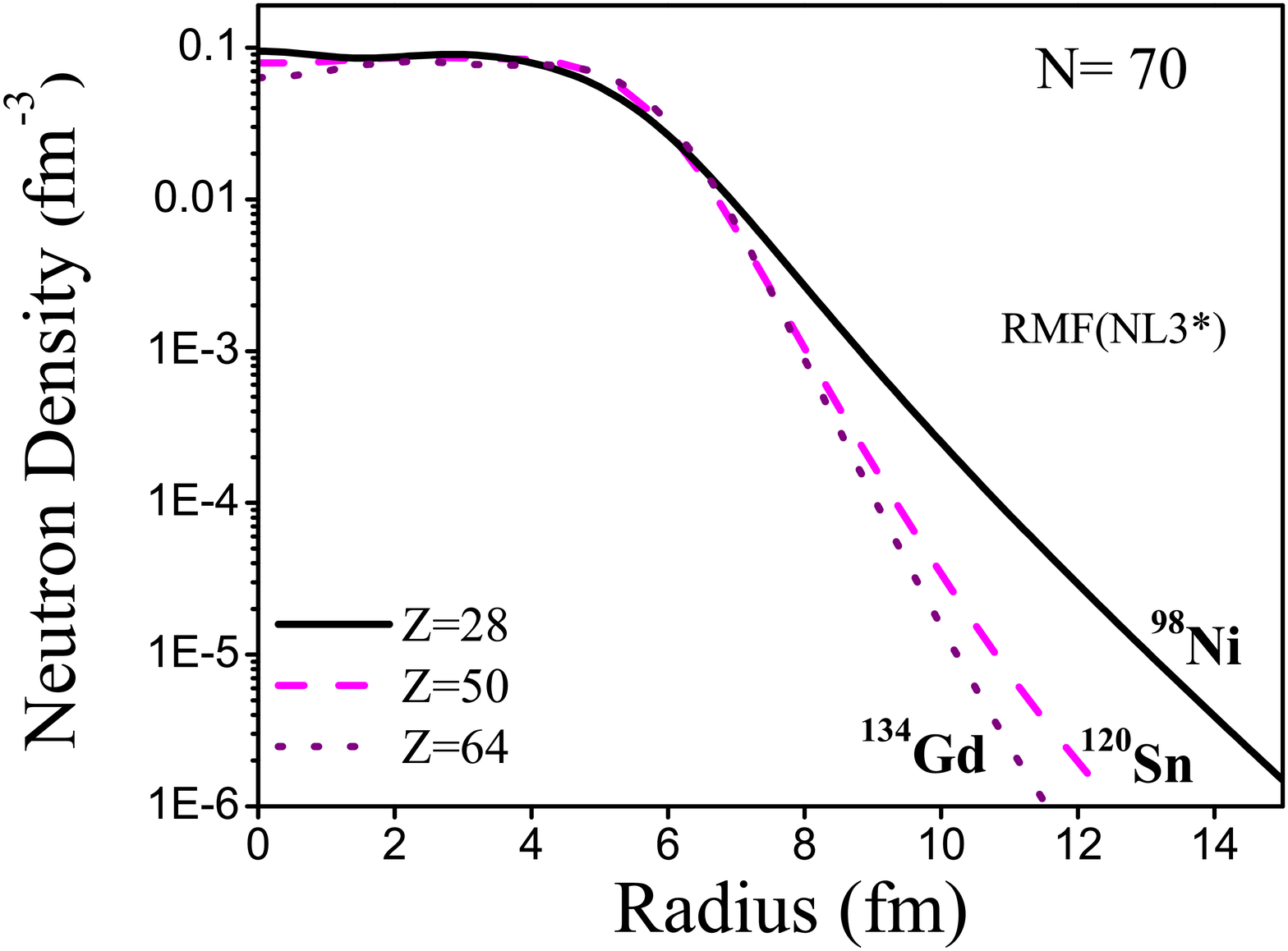}
\caption{(Colour online) Variation of neutron density for N$=$70 isotones with respect to radius.} \label{fig7}
\end{figure}

\begin{figure}[!htbp]
\centering
\includegraphics[width=0.6\textwidth]{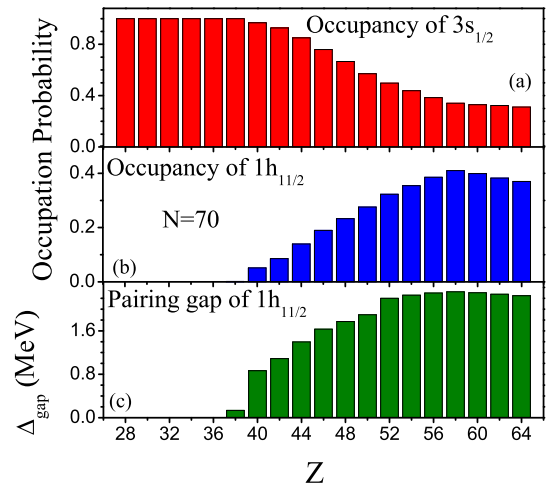}
\caption{(Colour online) Variation of (a) Occupancy of 3s$_{1/2}$, (b) Occupancy of 1h$_{11/2}$, and (c) Pairing gap of 1h$_{11/2}$ with respect to Z are shown.} \label{fig8}
\end{figure}

To investigate the magicity of N$=$70 in further details, we show the neutron density distribution of isotones of N$=$70 in Fig. \ref{fig7}. The sharp fall in the neutron density of isotones of N$=$70 shows its magic character. But the density distribution of $^{98}$Ni shows little more spread than other isotones although the energy gap (as seen in Fig. \ref{fig4}) between s.p. states (s-d and h$_{11/2}$) causing the shell closure at N$=$70 is maximum among the other N$=$70 isotones. To explain this discrepancy, we closely examine the neutron single-particle states along with their occupancy and pairing gap in Fig. \ref{fig8}. As is well known now \cite{terasaki,kaushik}, that if any low angular momentum state especially s-states ($\ell$$=$0) get occupied by the last filled particles near the Fermi level, then a large spatial extension of the density distribution is seen because of no centrifugal barrier of s-state. In case of $^{98}$Ni, the 3s$_{1/2}$ state get occupied fully by few of the last neutrons and hence the neutron density displays significant spatial extension due to zero centrifugal barrier.\par

For other isotones of N$=$70 ($^{120}$Sn and $^{134}$Gd) shown in Fig. \ref{fig7}, the occupancy of 3s$_{1/2}$ state reduces and that of the higher angular momentum state (1h$_{11/2}$) increases. The changing occupancies in 3s$_{1/2}$ and 1h$_{11/2}$ with increasing Z has been shown in a very systematic manner in Fig. \ref{fig8}(a) and (b). This figure shows the variation of occupancy of proton 3s$_{1/2}$ and 1h$_{11/2}$ along with the pairing gap of 1h$_{11/2}$. The occupancy of 3s$_{1/2}$ is maximum for $^{98}$Ni (Z$=$28) whereas 1h$_{11/2}$ does not occupy any particle (Fig. \ref{fig8}(a) and (b)). With increasing Z, in particular, after Z$=$40, the pairing gap of 1h$_{11/2}$ (Fig. \ref{fig8}(c)) starts increasing gradually which is a signature of the interaction with lower bound states. Such increased pairing gives rise to higher occupancy in 1h$_{11/2}$ and the occupancy in 3s$_{1/2}$ decreases gradually and becomes a minimum for $^{134}$Gd (Z$=$64) which is evident in Fig. \ref{fig8}(a). This figure explains the structural aspects of N$=$70 isotones due to deviation in the s.p. state spectrum. As Z increases, occupancy in 3s$_{1/2}$ decreases and that in 1h$_{11/2}$ increases. Occupancy in higher s.p. state results in more of pairing interaction. The centrifugal barrier for the last filled valence 1h$_{11/2}$ state being a higher $\ell$ ($=$5) state is higher which leads to smaller density distribution in (Z$=$40-64) nuclei. Therefore the density distribution curve (Fig. \ref{fig7}) is the outcome of the combined effect of occupancy in (3s$_{1/2}$,1h$_{11/2}$), pairing energy and the centrifugal barrier.\par

\begin{figure}[!htbp]
\centering
\includegraphics[width=0.6\textwidth]{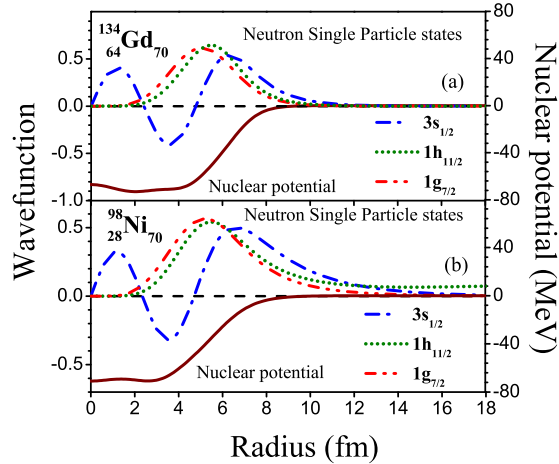}
\caption{(Colour online) The RMF potential energy (right scale) and wave-function (left scale) of 2d-3s states and 1h$_{11/2}$ for the case of $^{98}$Ni and $^{134}$Gd in N$=$70 isotones.} \label{fig9}
\end{figure}

To elaborate more on the difference in the density distribution of $^{134}$Gd and $^{98}$Ni, we have plotted wave-functions of 2d-3s states along with wave-function of 1h$_{11/2}$ in Fig. \ref{fig9}(a) and (b), respectively. The RMF potential energy which is a sum of the scalar and vector potentials is also shown for both the nuclei. A close watch of Fig. \ref{fig9} gives us a clear picture of 3s$_{1/2}$ state which is spread over outside the potential region for $^{98}$Ni causing the poorer overlap with the bound states near the Fermi surface and subsequently leading to a larger spatial extension of neutron density. On the other hand, for $^{134}$Gd, wave-function of both 3s$_{1/2}$ and 1h$_{11/2}$ states are confined within the potential range of about 8 fm. Therefore, N$=$70 strongly presents its candidature as a new neutron magic number with a unique nucleus $^{98}$Ni which is a nucleus with doubly magic character (see Fig. \ref{fig2}) and having a weakly bound or halo-like structure (see Fig. \ref{fig7}) at the same time which is reported here for the first time and is one of the highlights of this work.\par

On one side, as seen in the case of $^{98}$Ni that the occupancy of s-state (3s$_{1/2}$) ($\ell$$=$0) leads to weakly bound or halo-like structure (seen in Fig. \ref{fig7}) whereas on the other side, unoccupancy of s-state may lead to the "Bubble Structure". The depletion of central density comparative to its value at other radial distance leads to the phenomena of bubble-like structure. This phenomenon was recently observed in $^{34}$Si~\cite{nature} and studied by many theoretical works~\cite{duguet,todd,grasso,khan,wang,wang1,grasso1,yao1,li,schuetrumpf,wu} along with our recent work~\cite{saxena7,saxena8,saxenajpg} which have established $^{34}$Si as a best candidate showing bubble or central depletion in its charge density distribution. Encouraging with these recent studies, we have analyzed density distribution at the center for all the isotones of N$=$50, 70, 82 and 126 which may be influenced by occupancy in 3s$_{1/2}$ state.\par

\begin{figure}[!htbp]
\centering
\includegraphics[width=0.6\textwidth]{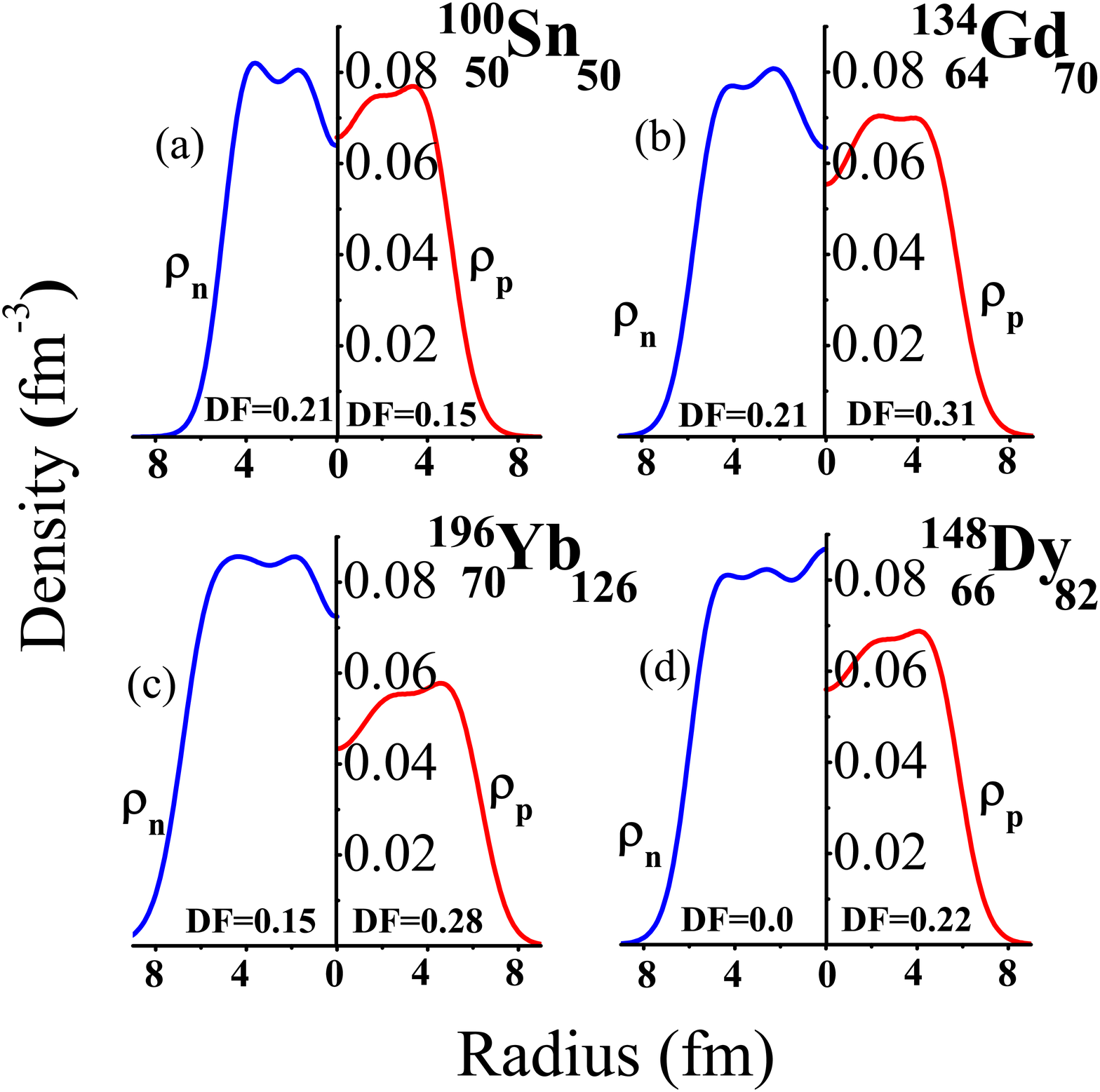}
\caption{(Colour online) Proton and neutron densities of $^{100}_{50}$Sn, $^{134}_{64}$Gd, $^{148}_{66}$Dy and $^{196}_{70}$Yb from N$=$50, 70, 82 and 126 isotonic chain, respectively.} \label{fig10}
\end{figure}

To search for the exotic phenomenon of bubble structure in 2d-3s shell nuclei, we have shown proton and neutron densities of some selected nuclei which are found with significant central depletion in Fig. \ref{fig10}. We identify the nuclei $^{100}_{50}$Sn, $^{134}_{64}$Gd, $^{148}_{66}$Dy and $^{196}_{70}$Yb as bubble nuclei among the N$=$50, 70, 82 and 126 isotonic chains.\par

To quantify the depletion of the central density, we have calculated depletion fraction (DF) ($(\rho_{max}$-$\rho_{c})/\rho_{max}$)~\cite{grasso,saxena7} which is also mentioned in the Fig. \ref{fig10}. The most remarkable observation from this figure is that the nuclei  $^{100}_{50}$Sn and $^{134}_{64}$Gd are showing the depletion in both the proton and neutron densities which predict the double bubble structure. These two nuclei along with $^{196}_{70}$Yb also qualify to be the candidates of doubly bubble nuclei. In $^{100}_{50}$Sn and $^{134}_{64}$Gd nuclei, the s.p. state 3s$_{1/2}$ remains empty for both protons and neutrons whereas for the case of $^{196}_{70}$Yb the proton depletion is due to empty 3s$_{1/2}$ and neutron depletion is due to empty 4s$_{1/2}$. In Fig. \ref{fig10}(d), we note that in case of N$=$82,  all the neurons keep 3s$_{1/2}$ completely filled and therefore no neutron depletion arises in $^{148}_{66}$Dy contrary to proton depletion in which Z$=$66 observes vacant 3s$_{1/2}$ proton state. The values of DF mentioned in the figure also explain this feature. From these chains of isotones, it is found that many of the nuclei possess the central depletion in their proton or neutron or both the densities.\par

\begin{figure}[!htbp]
\centering
\includegraphics[width=0.6\textwidth]{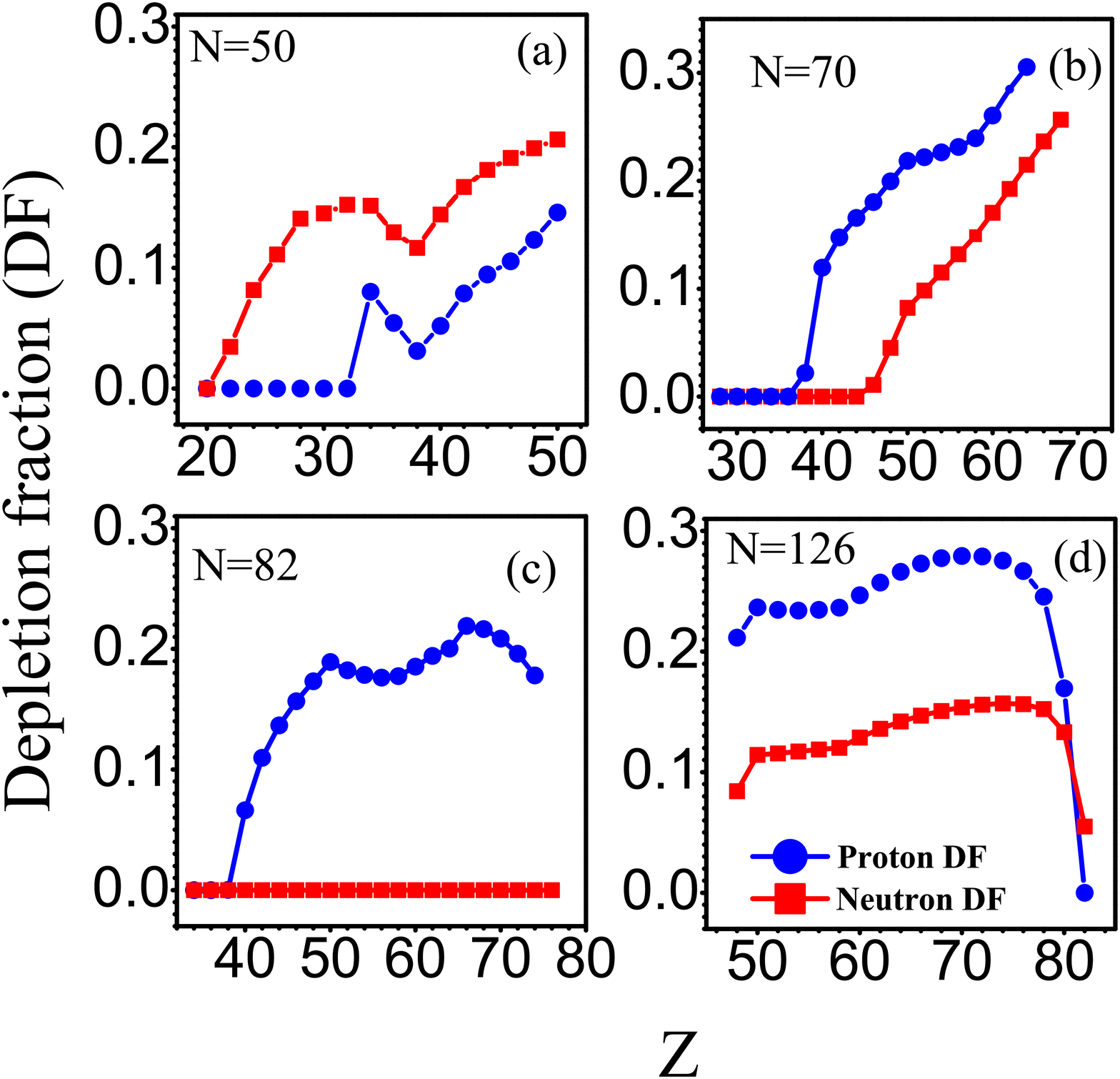}
\caption{(Colour online) Proton and neutron depletion fraction (DF) ($(\rho_{max}$-$\rho_{c})/\rho_{max}$) \cite{grasso} for N$=$50, 70, 82 and 126 isotonic chain.} \label{fig11}
\end{figure}

For a systematic study of both depletion, we have displayed proton and neutron depletion fraction (DF) for a full isotonic chain of N$=$50, 70, 82, and 126 in Fig. \ref{fig11}. Doubly bubble character is clearly observed for many of the isotones for N$=$50, 70, and 126 in Fig. \ref{fig11}(a), (b) and (d), respectively. In N$=$82 isotonic chains (Fig. \ref{fig11}(c)), zero neutron DF provides evidence of the contribution of 3s$_{1/2}$ state in bubble formation, which is completely occupied hence the central depletion is zero. One interesting fact which can be seen in Fig. \ref{fig11}(b), (c), and (d) is that DF increases abruptly (though there is only a hint) at magic number Z$=$40 and 50 as compared to their neighbourhood nuclei. In a similar manner, from Fig. \ref{fig11}(a) an abrupt increase in proton DF is found for Z$=$34 which might be correlated to new proton magicity in Z$=$34 \cite{stepp}. It is essential to note here that all these hints of magicity on the basis of DF are very preliminarily and need further investigation and more systematic study which is left for our more comprehensive subsequent work on bubble structure. On the other side, a fall in DF can also be envisioned from the figure of N$=$82 and 126 while reaching towards Z$=$82, due to filling of proton 3s$_{1/2}$ state.\par

\begin{figure}[!htbp]
\centering
\includegraphics[width=0.6\textwidth]{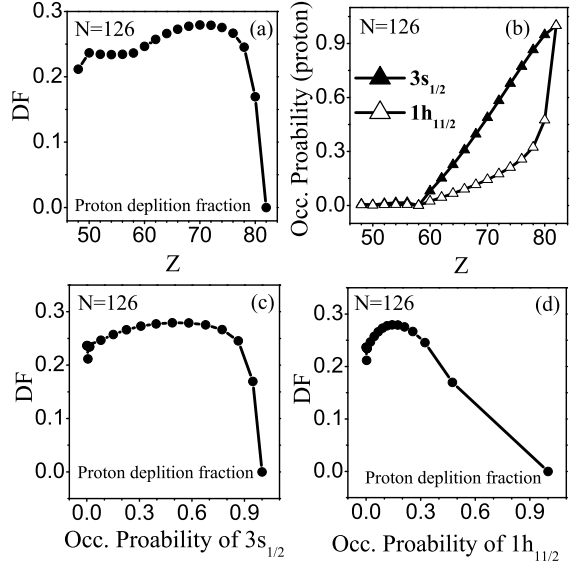}
\caption{(a) Proton depletion fraction (DF) ($(\rho_{max}$-$\rho_{c})/\rho_{max}$)~\cite{grasso}, (b) occupation probability of 3s$_{1/2}$ and 1h$_{11/2}$ states are shown for N$=$126 isotonic chain. Variation of DF with respect to occupation probability of (c) 3s$_{1/2}$ and (d) 1h$_{11/2}$ states is also shown.} \label{fig12}
\end{figure}

To elaborate more about the dependence of proton depletion on the occupancy of proton 3s$_{1/2}$ state, we have shown a few systematics of N$=$126 isotones in Fig. \ref{fig12}. Fig. \ref{fig12}(a) shows the variation of proton DF for full chain of N$=$126 isotones which is high for Z$=$50-78 and suddenly drops at Z$=$82 where the occupancy of 3s$_{1/2}$ is full. Fig. \ref{fig12}(b) depicts occupation probability of proton single-particle states 3s$_{1/2}$ and 1h$_{11/2}$  as a function of Z. As Z increases the occupancy of 3s$_{1/2}$ and 1h$_{11/2}$ states increase gradually. Although, it seems from Fig. \ref{fig12}(b) that the probability of occupying 1h$_{11/2}$ is always lesser than 3s$_{1/2}$ state but in terms of occupying the total number of particle the 1h$_{11/2}$ always has more particle than 3s$_{1/2}$ for Z$>$58. It is worthy to mention here that in addition to the necessary condition of unoccupied s-state for the bubble formation, the s-state should be surrounded by larger $\ell$ states which leads to weaker dynamical correlations consequently enhances bubble effect~\cite{saxena7}. The variation of DF in N$=$126 isotones indeed fulfills the above criteria and therefore appears with maximum proton DF rather than N$=$50, 70, and 82 isotones as can be seen from Fig. \ref{fig11}. The influence of occupancy of both 3s$_{1/2}$ and 1h$_{11/2}$ states on proton DF can be easily seen from Fig. \ref{fig12}(c) and (d). As soon as both the states get filled completely for Z$=$82, the depletion disappears. Hence, both states simultaneously determine central depletion in N$=$126 isotones. However, after a close watch the influence of 3s$_{1/2}$ state can be seen more dominating and wider as compared to 1h$_{11/2}$ state.\par

\begin{figure}[!htbp]
\centering
\includegraphics[width=0.6\textwidth]{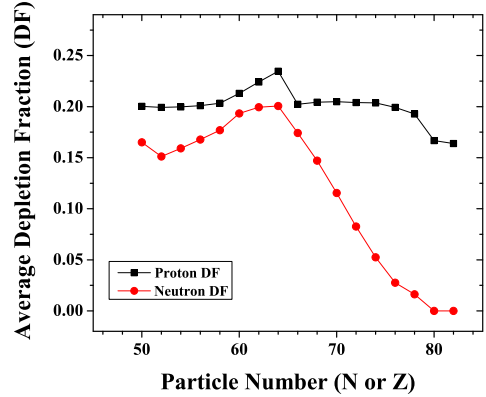}
\caption{(Colour online) Average proton and neutron depletion fraction (DF) for the considered nuclei in the range 50$\leq$Z(N)$\leq$82.} \label{fig13}
\end{figure}

While discussing the bubble phenomenon in medium or heavy mass region, the role of Coulomb repulsion becomes substantial. The impression of Coulomb repulsion on bubble formation has been discussed in $^{204,206}$Hg \cite{wu2018}. We also emphasized the role of Coulomb forces in superheavy region in our earlier works \cite{saxena7,saxena8}. To view the effect of Coulomb repulsion in this region of interest (50$\leq$Z(N)$\leq$82), we have calculated proton depletion fraction (proton DF) and also neutron depletion fraction (neutron DF) for all the considered isotopic and isotonic chains. The average DF is plotted in Fig. \ref{fig13} for proton and neutron depletion separately. For the same region (50$\leq$Z(N)$\leq$82), proton DF is always larger than neutron DF which is clearly an indicator of effect of Coulomb repulsion. A more detail analysis of effect of Coulomb repulsion on bubble structure would be reported in our subsequent works.

\section{Conclusions}
Various exotic features are probed in a less-known region which connects two conventional and strong magic numbers: Z(N)$=$ 50 and 82. This realm is influenced by the 2d-3s shell (sometimes referred to as second sd-shell) which leads to new magicity, the disappearance of conventional magicity, weakly bound or halo-like structure, and bubble structure. For this investigation, we have used relativistic mean-field approach along with the NL3* parameter to calculate deformation, binding energies, separation energies, pairing energies, single-particle energies as well as density distributions, etc. The results are compared with one more calculation using triaxially deformed Nilson Strutinsky model (NSM) and also FRDM, RCHB along with available experimental data and are found in excellent agreement. In this region, N$=$82 and Z$=$50, 82 are established as very strong magic nuclei whereas N$=$50 is found to disappear towards neutron drip-line for $^{70}$Ca. In addition to this, N$=$70 and Z$=$58 are found to bear witness to new magicity or sub-shell closure. Among isotones of N$=$70, $^{98}$Ni came across with unique behaviour as it is characterized by its double magicity and weakly bound halo-like structure, simultaneously. This weakly bound structure is observed due to occupancy in 3s$_{1/2}$ state: the state which also leads to central depletion in many considered nuclei, if empty. As a very important consequence, many double bubble nuclei are predicted with $^{100}_{50}$Sn, $^{134}_{64}$Gd and $^{196}_{70}$Yb as the most effective examples.
\section{Acknowledgement}
Authors G. Saxena and Mamta Aggarwal acknowledge the support provided by SERB (DST), Govt. of India under CRG/2019/001851 and WOS-A scheme, respectively.

\end{document}